# Investigation of Minerals Using Hyperspectral Satellite Imagery in Bangladesh

Nazmul Hasan[1Φ*], Kazi Mahmudul Hasan[2Φ], Md. Tahsinul Islam[3], Shahnewaz Siddique[4]

*Department of Electrical and Computer Engineering, North South University, Dhaka, Bangladesh*

[1]nazmul.hasan05@northsouth.edu, [2]kazi.hasan171@northsouth.edu, [3]tahsinul.islam@northsouth.edu, [4]shahnewaz.siddique@northsouth.edu

*Abstract*— Mineral identification using remote sensing technologies is becoming more dominant in this field since it saves time by demonstrating a more effective way for land resources survey. In such remote sensing technologies, hyperspectral remote sensing (HSRS) technology has increased gradually for its efficient manner. This technology is usually used from an airborne platform, i.e., satellite. Hence, satellite imagery remote sensing technology is now more capable of providing accuracy in mineral identification, and mapping. Hyperspectral satellite imagery can identify minerals more accurately compared to traditional technologies in remote sensing by constructing a complete reflectance of the spectrum from each pixel with its advanced imaging sensor. Bangladesh is a developing country with an area of 1,50,000 square kilometers located in Southeast Asia. Though it is a small country, it is enriched with several mineral resources through rivers, forests, hills, and the Bay of Bengal. In this study, hyperspectral imaging technology is employed on some major identical areas (Maheshkhali, Netrokona, Panchagarh, and Patuakhali) of Bangladesh to identify minerals there. As there are no studies done in Bangladesh using hyperspectral imaging yet, it is a good opportunity to explore the potentiality of HS imagery in this field. In this study, the FLAASH (Fast Line-of-sight Atmospheric Analysis) module with necessary parameter settings is used to filter the data, and finally, mineral identification is done by the spectral matched filtering method. Our investigation resulted in finding some potential minerals in those areas including Stariolite, Diasphore, Zircon, Alunite, Quartz, and so on. This indicates that there still is enormous potential for further exploration of minerals in Bangladesh by Hyperspectral Satellite Imagery.

## Declarations

The authors did not receive any financial support from any organization for the submitted work. All authors certify that they have no affiliations with or involvement in any organization or any competing interests to declare that are relevant to the content of this article.

Φ The authors are contributed equally.
*Corresponding author:
nazmul.hasan05@northsouth.edu

# Investigation of Minerals Using Hyperspectral Satellite Imagery in Bangladesh

*Abstract*— Mineral identification using remote sensing technologies is becoming more dominant in this field since it saves time by demonstrating a more effective way for land resources survey. In such remote sensing technologies, hyperspectral remote sensing (HSRS) technology has increased gradually for its efficient manner. This technology is usually used from an airborne platform, i.e., satellite. Hence, satellite imagery remote sensing technology is now more capable of providing accuracy in mineral identification, and mapping. Hyperspectral satellite imagery can identify minerals more accurately compared to traditional technologies in remote sensing by constructing a complete reflectance of the spectrum from each pixel with its advanced imaging sensor. Bangladesh is a developing country with an area of 1,50,000 square kilometers located in Southeast Asia. Though it is a small country, it is enriched with several mineral resources through rivers, forests, hills, and the Bay of Bengal. In this study, hyperspectral imaging technology is employed on some major identical areas (Maheshkhali, Netrokona, Panchagarh, and Patuakhali) of Bangladesh to identify minerals there. As there are no studies done in Bangladesh using hyperspectral imaging yet, it is a good opportunity to explore the potentiality of HS imagery in this field. In this study, the FLAASH (Fast Line-of-sight Atmospheric Analysis) module with necessary parameter settings is used to filter the data, and finally, mineral identification is done by the spectral matched filtering method. Our investigation resulted in finding some potential minerals in those areas including Stariolite, Diasphore, Zircon, Alunite, Quartz, and so on. This indicates that there still is enormous potential for further exploration of minerals in Bangladesh by Hyperspectral Satellite Imagery.

## I. Introduction

Sustainable economic development leveraging the UN's sustainable development goals (SDGs) for a developing country like Bangladesh is needed as a must to develop the strength of its mineral resources. The utilization of mineral resources has played a vital role in shaping the modern industrial world of civilization and is continuing to do so. The sustainable socio-economic infrastructure of any country indicates its natural resource wealth and technological strength in exploiting mineral resources for national development. Bangladesh is lagging far behind in mineral resources utilization compared to the developed countries due to the lack of adequate resources, technological abilities, and proper utilization.

Despite being a small country, Bangladesh is enriched with mineral resources such as natural gas, oil, coal, hard rock, limestone, white clay, glass sand, and mineral sand. Currently, the contribution of minerals to the economy of Bangladesh is mainly being satisfied using natural gas [1]. To achieve sustainable mineral development in comparison with the developed world, Bangladesh has to follow a couple of strategies to overcome its obstacles. Regarding this concern, spectral imaging technology is an option of remote sensing which can be practiced in such mineral identification in Bangladesh.

For spectral imaging, it can entail acquiring image data in both visible and non-visible bands at the same time. Capturing hundreds of wavelength bands for each pixel in an image can also be achieved [2]–[10]. Spectral imaging allows an extraction of additional information that the human eye fails to catch with its visible receptors for red, green, and blue. Spectral imaging is categorized into two groups, based on the number of bands and how narrow the bands are: (i) Multispectral Imaging and (ii) Hyperspectral Imaging. However, more advanced and reliable technology is Hyperspectral Imaging (HSI) which can identify minerals more accurately with comparatively large spectral information than other traditional methods as of its continuous manner with a narrow-wide range of bands availability [11]. In our focus of study, it has been directed towards identifying minerals from some identical zone of Bangladesh using the HS technology. It is an effective way for remote sensing applications considering its ability to discover richer information with the presence of greater spectral band data. To exercise such investigations in this study, imagery data from the USGS library for Earth Observing (EO-1) satellite's Hyperion sensor is used. Hyperion reflectance data after atmospheric correction remains consistent with the actual states [11], [12]. Hyperspectral imaging can provide a better continuous spectrum of objects and much more abundant information with corresponding electromagnetic waves [2], thus reliability with efficiency is gained from such identification.

Hyperspectral imaging is a sophisticated technique also known as Reflectance imaging spectroscopy since the images depict the reflectance values corresponding to a wide range of wavelengths. Using a major part of the electromagnetic spectrum, the hyperspectral Hyperion sensor can look at objects through a total of 242 bands [13], [14]. Certain objects in the electromagnetic spectrum leave distinctive 'fingerprints', known as spectral signatures, and these spectral signatures allow the materials that make up a scanned object to be identified [5], [6], [9], [11], [15].

According to our systematic review, no approach to mineral identification using hyperspectral image-based remote sensing has yet been accomplished, but similar mineral identification has been attempted through the studies [10], [16]–[18] in Bangladesh using typical physical material characterization processes and multispectral imaging technology. Hence, with the motivation of applying HS imaging technology to the mineralogy sector in Bangladesh, minerals identification over many regions of the country is being investigated in this study. These HSI investigated results benchmark the novelty of employing the imagery in such mineral explorations in Bangladesh.

The rest of the paper is organized as follows: Section II provides a brief overview of the extant works of literature to establish the concepts and terminologies of HSI technology and its future perspective for Bangladesh of it. Section III (Methodology) explains how we approached this study by briefing the flow of methods for this investigation. Section IV presents a discussion on mineral identification illustrating the findings from the selected regions. Section V suggests some future research scopes on this technology, where concluding remarks are provided in Section VI.

II. PAST WORK

An extensive review of the existing works of Hyperspectral Imaging technology applications in the mineralogy domain and the recent works based on the exploration of minerals in Bangladesh were overviewed. In order to situate the work in the literature, the past works are divided into two categories- Hyperspectral Imagery Technology and Exploration scopes of minerals in Bangladesh using HSI.

*A. Hyperspectral Imagery Technology*

Hyperspectral imaging (HSI) is an advanced technology of remote sensing for mineral detection processes. HS Imaging (Spectroscopy) can reliably measure the absorption characteristics of narrow bandwidth with appropriate adjacent spectral samples where a sampling limitation of <25 nm is utilized for the identification and classification of mineral features. As a single pixel may comprise a mixture of spectral characteristics, a Hyperspectral sensor can collect and make the picture more informative following its moving path. Hyperspectral Images can record a total of 242 bands (order of 0.01 µm) to construct a complete reflectance spectrum of every pixel in the scene whereas Multispectral Images usually contain 3-10 bands. Hence, Hyperspectral Imaging shows a continuous spectrum profile. Reflectance values of a hyperspectral image depend on meteorological characteristics, material and chemical compositions, and physical structure, which RGB and Multispectral images cannot visualize and acquire [3], [9], [11], [13]–[15], [19], [20].

In the study [13], the authors demonstrated an experimental investigation following the basic principles of HS remote sensing in mineral exploration. The HS data taken from the Hyperion (EO-1) sensor is more consistent in processing the actual information being projected in the study [11] by introducing FLAASH analysis. The authors demonstrated the steps of processing with visualizing results to make reliable the result compared with USGS spectral library data. A new diversity in HS imaging technology using the features of Long Wave Infrared data (LWIR) was demonstrated in the study [2], [3] with the advantages of mineral classification by utilizing optical and thermal HS data. Another improved algorithm was exhibited through the study [4], [5] on mineral mapping using the HS technology with spectral information divergence, and the distributive concept of probability. In the study [6], the authors used linear spectral unmixing with hyperspectral VNIR-SWIR drill-core scanning to map vein-hosted mineralization. Furthermore, in [7] authors performed their investigations on characterization after detecting oil slicks by combining active fluorescence laser technology with hyperspectral imagery (HSI) techniques on SAR. In [8], the authors presented a comparative view on interpolations from traditional analysis to HS analysis by airborne remote sensing. Applying this new technology of remote sensing, the process of mineral identification and mapping of altered zones has come up with satisfactory conclusions in different countries as exhibited in the literature [9], [11], [12], [15], [20]–[23]. Moreover, the studies [24], [25] have presented more advanced ways of RS by implementing the resonant microwave cavity concept in hyperspectral SAR imaging. Beyond the mineral explorations, the authors employed the Hyperspectral Imaging technique in [26] to corroborate its rapid ability as an application in food security.

So, there have been several comprehensive reviews of the unwavering advances in remote sensing and mineral exploration made possible by Hyperspectral Imagery. As the science of mineralogy advances rapidly, we must maintain this trend in order to assess our progress toward achieving long-term sustainability.

*B. HSI Scopes in Exploration of minerals in Bangladesh*

Bangladesh is a naturally wealthy country in a subtropical climate. The country's geographical conditions favor the availability of many natural minerals. However, due to a lack of technological advancement, such minerals have yet to be explored. Although extensive Hyperspectral Imaging-based projects have not yet been undertaken in Bangladesh, several experimental works on mineralogy have been conducted using other methods such as manual investigation, material characterization, and multispectral imaging (MSI).

Recently, some researchers investigated and discovered the most valuable uranium deposit in the country land, which they presented in their studies [17], [18], [27], [28] using physical characterization methods such as XRD, XRF, and SEM. Another study analyzed bulk beach sands and individual mineral fractions for environmental radioactivity along 120 km of Bangladesh's Cox's Bazar coastline [16]. The research [10] focuses on the identification of minerals using Multispectral Imaging techniques, which have been found to be useful for mineral exploration in coastal and remote areas of Bangladesh where hand-on investigation appears to be unsuitable and costly.

HSI could solve long-term problems in Bangladesh's mineral sector. As evidenced by the literature, mineral exploration and altered zone mapping using hyperspectral imaging technology are lacking. This study contributes to the exploration of minerals in major regions in Bangladesh to congregate mineral resource opportunities. Using hyperspectral imagery (HSI), we tried to establish a link between the country's mineral resources and sustainable development opportunities.

III. STUDY AREA AND METHODOLOGY

Bangladesh is mainly riverine. It's the eastern two-thirds of the Ganges and Brahmaputra delta plain north of the Bay of Bengal. Madhupur Tract and Barind Tract are two small areas of higher land in the north-center and north-west made of old alluvium. Along the eastern border are steep, folded Tertiary hill ranges. The basement rocks under central and southern Bangladesh are downwarped by Cretaceous sediments, mostly carbonate. Bangladesh's surface rocks and stratigraphy date to the Cenozoic, but the Oligocene is poorly preserved. Drilling and mineral prospecting has revealed deeper rock units that record Bangladesh's pre-Cenozoic geologic history [29]–[31].

In this study, it is penned through a comprehensive literature study focusing on three aspects: 1. Insights of Hyperspectral (HS) Imagery, 2. Related works on HS approach to mineral identification, and 3. The current progress of mineral exploration research in Bangladesh. Following the past works [10], [16]–[18] on mineral exploration in Bangladesh, Hyperspectral imagery is adopted for this process in our investigation. This becomes more convenient as the country advances technologically, as evidenced by the presence of satellites in space. The hyperspectral imagery analysis for our study has followed its scientific analysis framework.

Hyperspectral imagery technology is employed in this investigation to exercise mineral identification on some identical zones of the Cox's Bazar area, the northwestern area, and the middle portion of the country. Hence, a comparative study is conducted using satellite hyperspectral imaging technology to identify minerals in Bangladesh.

This investigation is carried out using an easy-to-use Environment for Visualizing Images (ENVI) tool, which is integrated with advanced spectral image processing and validated by geospatial analysis technology. With its distinctively diverse features, the tool enables the establishment of spectral analysis with extensive information from images, as well as scientific mapping methods for imagery analysis in mineral identification. The tool is free to get, and the open-source resources made us more tempted to use it for our investigation. These investigations are carried out following ways using the ENVI tool as described as follows-

A. DATA ACQUISITION

In our study, the novel method for mineral classifications in Bangladesh using remote sensing technology is presented following the concept of mathematics, machine learning, and signal processing as well as data engineering on a hyperspectral image of the considered regions which are obtained by NASA's Earth Observing-1 (EO-1) satellite's Hyperion sensor. The United States Geological Survey (USGS) is the trusted institution for collecting such spectroscopic remote sensing datasets across the world, where data is being stored at their Earth Explorer (https://earthexplorer.usgs.gov/) online tool.

The desired region is obtained through the earth explorer tool with selecting parameters for EO-1 Hyperion sensor data appearing with the captured images as shown in figure (1). For this investigation in our study process, we have used the EO1H1360452003332110PZ_MTL_L1GST dataset obtained from the EO-1 satellite.

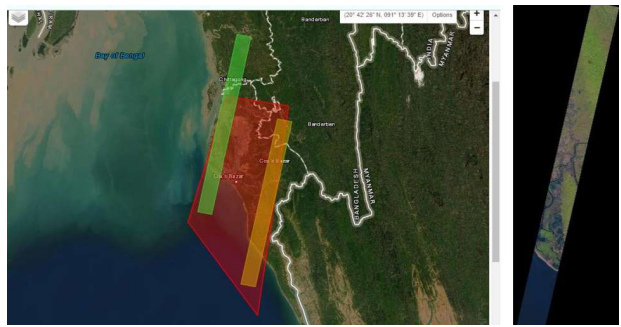

*Figure 1. The interface of data collection from USGS Earth Explorer, and a true-color version of the hyperspectral image.*

The Level-1 GST (L1GST) product is used in our experiment since in this particular product of data all radiometric and geometric corrections were performed by the USGS with a 90-meter digital elevation model. The geographical coordinates of the image are as:

Upper left (N-W) corner: Lat: 22°03'26.21"N Long: 91°58'28.16"E

Lower left (N-E) corner: Lat: 22°02'31.74"N Long: 92°02'48.81"E

Upper right (S-W) corner: Lat: 21°16'25.97"N Long: 91°47'16.95"E

Lower right (S-E) corner: Lat: 21°15'31.73"N Long: 91°51'36.24"E

B. DATA PRE-PROCESSING

Data preprocessing is required in order to apply any required algorithm to the highest competency on a more qualitative data set. Among the 242 bands of its definition, the acquired image contains some uncalibrated bands. As a result, an algorithm is used to remove these bad bands, resulting in a subset of this image on the relevant region of the image above. To obtain the only interested portion of the image for further classification, a region of interest (ROI) tool was used. ENVI performs a real-time orthorectification for its ROI tool using two different methods: one is creating ROI from the geometry selection and another is creating an ROI from the pixel information. For our investigation, we have performed using the geometry selection method to get our interested region. In order to exclude some redundant regions from the ROI proceeded image including the water vapor region, overlap region, and non-illuminated region, the bad bands are removed by applying the spectral subset algorithm and a subset of the actual metadata is shown in figure (2).

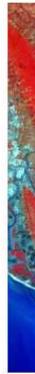

Figure 2. ROI processed image after bad band removal

There are many ways in which land cover impacts the environment, including water quality, the hydrology of watersheds and the diversity of species composition in the area as well as the climate and carbon storage which in turn produces atmospheric interferences. Atmospheric interferences can affect the spectral signature results for inconsistencies in radiance values and create interruptions to obtaining a good approximation of the result. Hence, atmospheric correction is a critical step in the processing of cloud-free atmospheric data. Properties such as the volume of water vapor, aerosol delivery, and scene visibility must be understood to compensate for the atmospheric impact. To perform the atmospheric correction, the Fast Line-of-sight Atmospheric Analysis of Hypercubes (FLAASH) module of the ENVI package is used to retrieve spectral reflectance from images of hyperspectral radiance. Details of the ENVI FLAASH algorithm are described in the study [32]. A radiometrically calibrated radiance image in the band-interleaved-by-line (BIL) format is used as the input image for FLAASH [12].

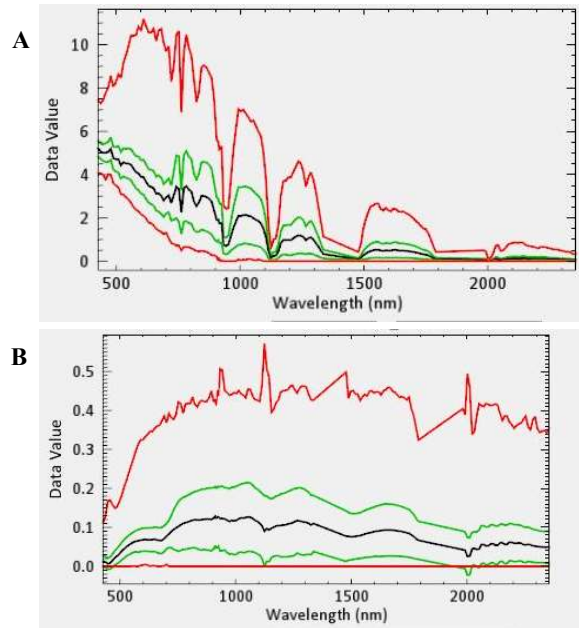

*Figure 3*. Conversion of Radiance values to Reflectance applying FLAASH on ROI: (A) Radiometric calibrated radiance profile, (B) Refined reflectance data after atmospheric correction optimized by band math.

The radiometric correction was performed on the ROI image data to obtain the radiance profile and then atmospheric correction was also employed to extract reflectance signature data from the image as shown in figure (3). In figure (3A), the radiance data with respect to the wavelength is depicted after the bad band removal process is done. The resultant outcome was employed through FLAASH to perform atmospheric correction that results in a reflectance profile as depicted in figure (3B). The mean data values are standardized for resampling and characterization advantages utilizing the band math optimization. In order to lessen the redundant effect of environmental factors, a standardization process was performed using the principal component analysis (PCA) algorithm [33]. Following this algorithm, the mean of spectral data of each pixel is subtracted from the original data to remove any kind of biases in pixels and normalizes concerning a standard deviation that is performed to minimize the effect of dissipated power variation among the spectral bands to each pixel. These processes are outcomes with the source reflectance signature as demonstrated in figure-(3B).

*C. DATA PROCESSING TO SPECTRAL ANALYSIS*

The atmospheric corrected preprocessed data is further employed for spectral analysis. In this process, filtering techniques are applied to the data image to extract mineral features from the image. The analysis framework is designed, integrated, and coordinated into the ENVI Spectral Hourglass Wizard as projected in figure (4), which guides the step-by-step processing flow to find and map image spectral endmembers from hyperspectral data and incorporates the K-means classification algorithm. This is a notable way to classify a given data set using the cluster means and assigning data elements to the closest cluster uses an iterative purity process where the inputs are the number of clusters and data set to produce a result. The details of the algorithm are described in [4], [5].

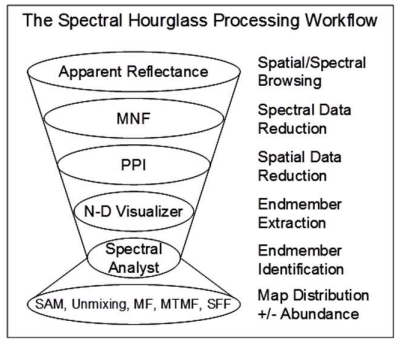

*Figure 4.* Analysis framework in ENVI software to extract mineral information [L3 Harris Geospatial, 2021]

The hourglass processing drift utilizes the spectrally over- definite nature of hyperspectral information to locate the most spectrally unadulterated, or unique pixels (called endmembers) into the dataset and to map their areas and subpixel abundance. This analyzing stream starts with reflectance input information, which can generate spectrally and spatially subsets of the data, envisioning the information in the n-D space visualizer, and bunch the purest pixels into endmembers. Mapping of the distribution and abundance of the endmembers, ENVI's Spectral Analyst is used to distinguish the endmembers, and audit the mapping results.

D.  MINIMUM NOISE FRACTION (MNF) TRANSFORMATION

The minimum noise fraction (MNF) transformation technique is used to separate the noise from the vast amount of data generated from the continuous spectrum of hyperspectral images and thus reduce data dimensionality using a linear transformation technique PCA. Methods exist for applying the MNF transform to eliminate noise from the data by performing a forward transformation to evaluate the bands which contain coherent images (by analyzing images and eigenvalues) and using a spectral subset to run an inverse transform to include only the good bands [34].

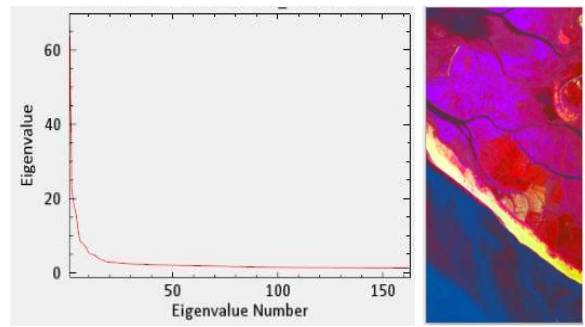

Figure 5. MNF transformed rescaled reflectance data.

For our particular data, an MNF transformation was performed on the FLAASH corrected image data, in which the first 48 eigenvectors of the MNF transformation contain coherent information which can be used for further processing as projected from the MNF transformed image data in figure (5).

E.  PIXEL PURITY INDEX (PPI)

Spectrally pure each pixel (Endmembers) in the hyperspectral image contains only features for a spectrally unique material. Pixel purity index (PPI) is typically performed on MNF data that has been reduced to coherent images to find the extreme pixels. By finding the purest pixels relative to other pixels in our image we can do subsequent analysis in selecting our end members. The result of the PPI routine is an image where the value of each pixel corresponds to the number of times it was identified as a pure pixel during all of the PPI iterations. These resultant pixels are excellent candidates for end-member selection that can be used in subsequent processing to the n-D visualizer. Spectrally pure, distinctive materials that occur in the scene are endmembers.

Authors have presented an improved algorithm of the PPI process in the study [35] introducing the advantages of virtual dimensionality (VD) to estimate the number of endmembers required as well as the studies [36]–[38] also have presented the PPI iteration process in details.

PPI iteration threshold value is given at an optimum level of 2.5 through the dialogue appeared by the ENVI PPI module. In figure (6), it exhibits the desired outcome plot from the iteration performed to get the best spectral pixels which are clustered to about 6200 pixels, and we examined this data for 10000 iteration times to get a better outcome.

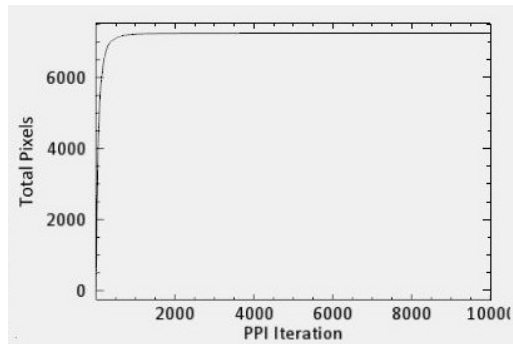

*Figure 6. Iteration plot of Pixel Purity Index process.*

*F. n-D VISUALIZER AND DERIVING ENDMEMBERS*

The filtered image data obtained from PPI analysis are exhibited by the n-D visualizer to find endmembers by locating and clustering the purest pixels in the n-dimensional space created by MNF. In an n-dimensional scatter graph, where n number of bands simply points out the spectral reflectance values for a given pixel in each band. In such n-dimensional space, the direction of each eigenvector distinguishes the spectral behavior of the particular pixel, which is considered as a basis of classification since the magnitudes vary by materials' physical factors. A detailed hypothesis is described on the geometric performing process of the n-D visualizer in the study manual [39]. A wide range of algorithms in controlling and mechanism of endmember selection is demonstrated in [38], [40]–[47] including Ant-Colony Optimization (ACO), Particle Swarm Optimization (PSO), and Random N-Finder.

When pixel data is plotted in a scatter plot with image bands as plot axes, the spectrally purest pixels always occur in the data cloud corners, while spectrally mixed pixels always occur on the inside of the data cloud. From our investigation, we have come up with a data cloud from the performed PPI data as shown in figure 7(A). The n-D visualizer selected bands of 48 classes were grouped as a means of purest pixels and the spectral reflectance of the group was demonstrated as endmember spectra in figure 7(B).

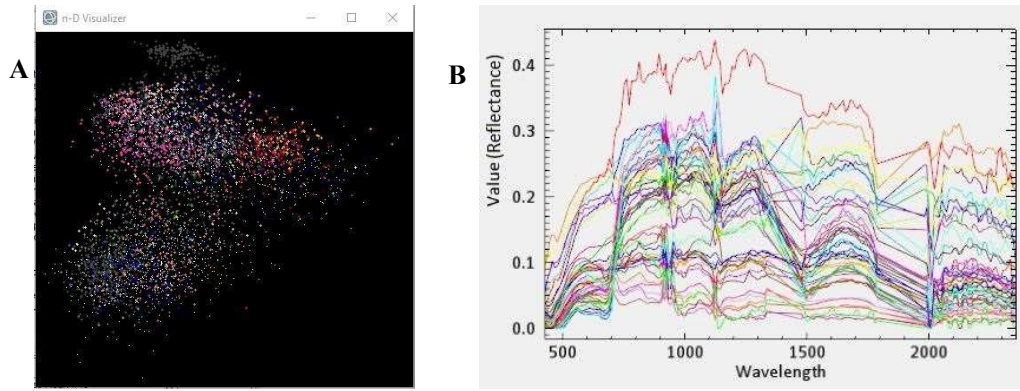

*Figure 7. n-D visualizer scatter (data cloud) plot (A) and visualization of Endmember spectra (B)*

The resulting endmember spectra were then employed in the spectral resampling method for further analysis to identify the minerals through field spectra comparison with USGS spectral library data, as projected in the following section.

IV. MINERAL IDENTIFICATION RESULTS

In our study, the endmember spectral signature data is compared to the USGS mineral library's spectral data which concludes towards a result. In the study [19], [22], [45], [49], a correlation analysis based on geochemical identity is utilized in the identification process, which can be useful for such analysis. The results are generated by comparing spectral similarity using the ENVI tool's Spectral Analyst module. A probability-based approach to spatial distribution is used to identify the reference spectral signatures for each class.

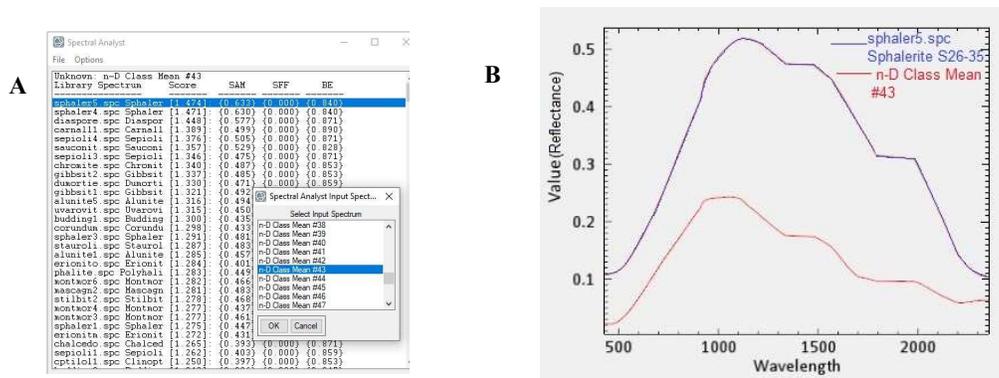

*Figure 8. Changing the n-D class mean value to 43 (A) and using Spectral Analyst, the known and unknown spectra matching (B)*

Using the Spectral Analyst to identify materials based on their spectral properties. The Spectral Analyst ranks the match of an unknown spectrum to the materials in a spectral library using ENVI techniques like Binary Encoding, Spectral Angle Mapper, and Spectral Feature Fitting. The Spectral Analyst produces a ranked or weighted score for each of the materials in the input spectral library. The highest score indicates the closest match and higher confidence in the spectral similarity. Similar materials may have relatively high scores, but unrelated materials should have low scores. Following this method of unmixing, in figure (8), it appears with the resultant unknown spectral signature when each eigenvector classified in n-D mean class values is changed, which is compared with the known laboratory spectral signatures accumulated from the USGS Spectral Library to specify. However, in comparison with the known spectral signature, the unknown spectral signature from the n-D class mean value of 43 exhibits the presence of Sphalerite composite as depicted in figure 8(B), while figure (9) shows the existence of Staurolite in our interested region (ROI) with the n-D class mean value of 47.

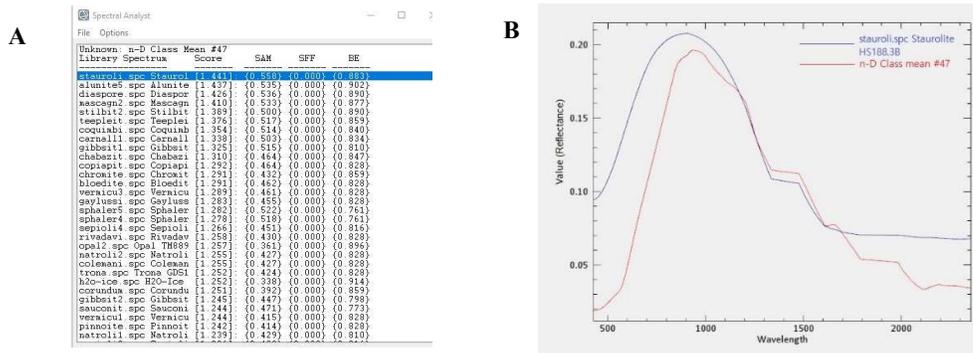

*Figure 9. Spectral Signature match filtering for n-D class mean value of 47 (A) and matched signature for Staurolite (B)*

We are examined for a partial portion of a large number of eigenvectors. In our study, results from our investigation are presented with identifying some minerals as shown in the supplementary figure S1-S4 with the most probable existence. Spectral Angle Mapper (SAM), a physical classification algorithm is used to compare the angle between the endmember spectrum vector and each pixel vector in n-D space resulting in finding the spectral similarity between the two endmember spectra [50]–[52]. MTMF (mixed tune matched filtering) algorithm maximizes a known endmembers' response while suppressing the background response [23]. As a result, the MTMF algorithm is used to perform Matched Filtering (MF), which limits the number of false positives because pixels with a high infeasibility are more likely to be MF false positives. Pixels that have been correctly mapped will have an MF score that is greater than around zero background distribution and a low infeasibility value. MTMF, on the other hand, necessitates performing the Minimum Noise Fraction (MNF) transform on the input file with unit variance noise, which was done using the ENVI hourglass wizard. In this case, the MNF transform was used to remove noise from data by performing a forward transform, determining which bands contain the coherent images (by inspecting the images and eigenvalues), and then running an inverse transform using a spectral subset to include only the good bands, or smoothing the noisy bands. However, figure 10(A) exhibits the mapping view by accumulating the spectral data for each class with its eigenvector properties acquired from the SAM whereas figure 10(B) illustrates the statistical information on the contribution of each band class with their pixel count.

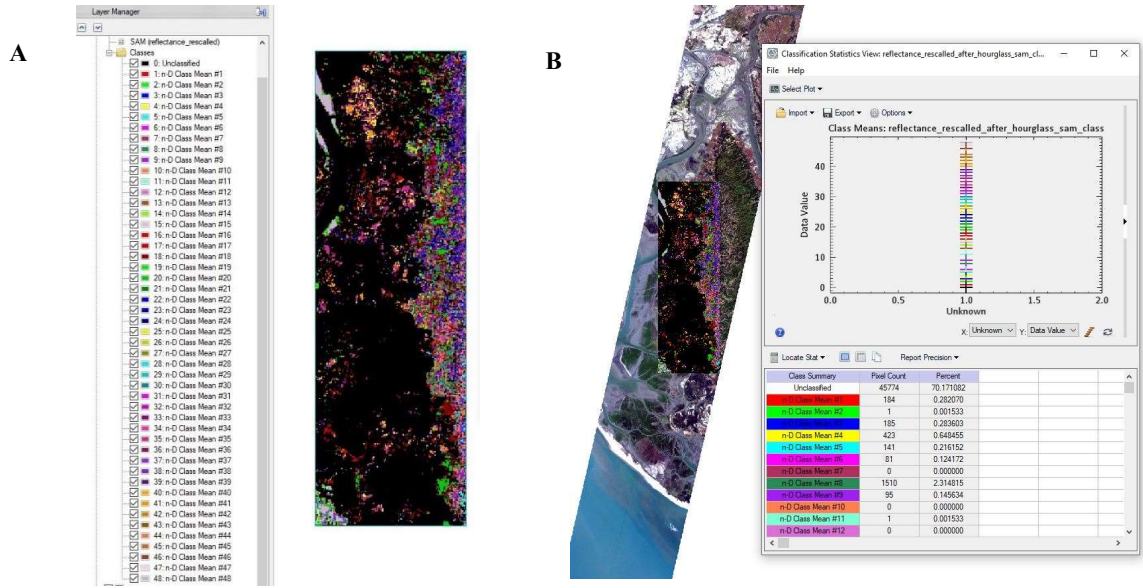

*Figure 10. (A) SAM viewed mapping for the 48 class mean bans from eigen-member spectra (B) Altered mapping from MTMF and percentage contribution of each mean class.*

The obtained results were validated by referring to the study [10] which used Multispectral Imagery to explore mineral deposits in Bangladesh's coastal areas and reported weight percentage data of light and heavy materials for the Moheshkhali region. In which, the most abundant minerals found in this studied area are Ilmenite (6.11 percent), magnetite (2.61 percent), zircon (1.22 percent), and staurolite (0.57 percent), among others and this concentration indicates that the region is more minerally enriched than another region of the country. In our study, we found the presence of Zircon, Staurolite, Opal, Corundum, Diaspore, and Pyrite as a high probable match from the spectral signatures for the Moheshkhali region as listed in Table-1, which validates the results primarily according to the findings of mineral elements in the prior study [10].

A summarized list of the findings corresponding to the resultant figures is listed in table 1 as-

*Table 1. n-D mean class varied explored minerals as high probability.*

| MOHESHKHALI REGION (South-East) | | |
|---|---|---|
| Figure | n-D mean class | Matched element |
| S1(A) | 2 | Opal |
| S1(B) | 5 | Corundum |
| S1(C) | 35 | Zircon |
| S1(D) | 38 | Diaspore |
| S1(E) | 45 | Pyrite |
| S1(F) | 47 | Staurolite |
| S1(G) | 48 | Alunite |
| NETROKONA REGION (North-East) | | |
| Figure | n-D mean class | Matched element |
| S2(A) | 26 | Alunite |

| Figure | n-D mean class | Matched element |
|---|---|---|
| S2(B) | 29 | Carnalite |
| S2(C) | 40 | Gibbsite |
| S2(D) | 4 | Mascagnite |
| S2(E) | 6 | Zircon |
| S2(F) | 15 | Rivadavite |
| S2(G) | 19 | Vemiculite |
| S2(H) | 22 | Sepiolite |
| S2(I) | 31 | Sauconite |
| S2(J) | 27 | Stilbite |
| S2(K) | 48 | Montmorillonite |

**PANCHAGARH REGION (North-West)**

| Figure | n-D mean class | Matched element |
|---|---|---|
| S3(A) | 6 | Sepiolite |
| S3(B) | 7 | Richterite |
| S3(C) | 21 | Chromite |
| S3(D) | 23 | Richterite |
| S3(E) | 24 | Kaolin/Smect |
| S3(F) | 26 | Nontronite |
| S3(G) | 27 | Buddingtonite |
| S3(H) | 30 | Zircon |
| S3(I) | 41 | Kaolin/Smect |

**PATUAKHALI REGION (South-West)**

| Figure | n-D mean class | Matched element |
|---|---|---|
| S4(A) | 24 | Alunite |
| S4(B) | 24 | Bloedite |
| S4(C) | 24 | Chabazite |
| S4(D) | 6 | Clinoptilolite |
| S4(E) | 4 | Coquimbite |
| S4(F) | 16 | Erionite+Merlinoit |
| S4(G) | 33 | Gibbsite |
| S4(H) | 21 | Kaolin/Smect |
| S4(I) | 33 | Montmorillonite |
| S4(J) | 21 | Nontronite |
| S4(K) | 1 | Quartz |
| S4(L) | 9 | Palygorskite |
| S4(M) | 19 | Sauconite |
| S4(N) | 9 | Sepiolite |
| S4(O) | 24 | Stilbite |
| S4(P) | 4 | Vermiculite |
| S4(Q) | 28 | Vesuvianite |

## V. FUTURE SCOPES

In future research, the accuracy of mineral mapping can be enhanced from the perspective of sub-pixel by endmember/end-element extraction and abundance inversion. Therefore, in addition to applying vein detection processing, automating endmember extraction with the implementation of the protocol to display the alteration view would allow for better and faster 3D modeling with identification. However, improvements in mapping and identification can be done using Gaussian convolution sigma parameter variation with improved kernel means (K-means) algorithm. Also, the improvement of the endmember extraction and unmixing methods could facilitate the linear feature extraction process with the application of machine learning using ENVI programming utility. The optical and thermal characteristics of the hyperspectral image can be utilized in more advanced processing. An improved SCM algorithm can be applied over the SAM process to make the optimized process of Hyperspectral Imagery.

## VI. CONCLUSION

In mineral prospecting and discovery, geological remote sensing using spaceborne technology plays an important role. The fundamental concept of remote sensing for deposit discovery is that various minerals in such remote sensed images have carried specific spectral signatures corresponding to their own composition variations. Compared with conventional techniques, hyperspectral remote sensing technology may identify minerals to create a continuous reflectance spectrum of every pixel in view.

From other prior exploration studies, most of the coastal belt of Bangladesh is shown to be enriched with valuable minerals including Zircon, Ilmenite, Magnetite, and so on. In several recent studies, uranium is estimated to have a high deposit volume in the South-Eastern Islands of the country. This study, therefore, demonstrates the feasibility of using Hyperion's hyperspectral (HS) image data (EO-1 satellite) to classify the altered mineral deposits following ENVI's remote sensing analysis processes.

In this study, the visual inspection has projected a good estimation yet, which demands more exploration in applying this advanced HS imaging technique in mineral detection. In our study, it is covered some identical areas of Bangladesh, but it raises the necessity to exercise on more areas with the availability of hyperspectral data to explore the resources. Exploration will make the thought and way to utilize mineral resources in building a strong and sustainable economy.